\documentclass[5p,twocolumn]{elsarticle}
\pdfoutput=1

\usepackage{lineno,hyperref}
\usepackage{amssymb} 
\usepackage{siunitx}
\usepackage{amsmath}
\usepackage{caption}
\usepackage{subcaption}
\usepackage{epstopdf}
\usepackage[utf8]{inputenc}

\journal{Nuclear Instrumentation and Methods in Physics Research A}

\newcommand{\matr}[1]{\mathbf{#1}}

\begin{document}

\begin{frontmatter}

\title{Longitudinal phase space reconstruction simulation studies using a novel X-band transverse deflecting structure at the SINBAD facility at DESY}

\author[DESY,UniHamburg]{Daniel Marx\corref{mycorrespondingauthor}}
\cortext[mycorrespondingauthor]{Corresponding author}
\ead{daniel.marx@desy.de}

\author[DESY]{Ralph Assmann}

\author[PSI]{Paolo Craievich}

\author[DESY]{Ulrich Dorda}

\author[CERN]{Alexej Grudiev}

\author[DESY]{Barbara Marchetti}

\address[DESY]{DESY, Hamburg, Germany}
\address[UniHamburg]{Universität Hamburg, Germany}
\address[PSI]{Paul Scherrer Institut, Villigen PSI, Switzerland}
\address[CERN]{CERN, Geneva, Switzerland}

\begin{abstract}
A transverse deflecting structure (TDS) is a well-known device for characterizing the longitudinal properties of an electron bunch in a linear accelerator.
The standard use of such a cavity involves streaking the bunch along a transverse axis and analysing the image on a screen downstream to find the bunch length and slice properties along the other transverse axis.
A novel X-band deflecting structure, which will allow the polarization of the deflecting field to be adjusted, is currently being designed as part of a collaboration between CERN, DESY and PSI.
This new design will allow bunches to be streaked at any transverse angle within the cavity, which will open up the possibility of new measurement techniques, which could be combined to characterize the 6D phase space distribution of bunches.
In this paper, a method is presented for reconstructing the longitudinal phase space distribution of bunches by using the TDS in combination with a dipole. 
Simulations of this technique for the SINBAD-ARES beamline are presented and the key limitations related to temporal resolution and induced energy spread are discussed.
\end{abstract}


\end{frontmatter}

\section{Introduction}
The ARES linac, currently under construction at DESY, will form an integral part of the SINBAD R\&D facility~\cite{EAACSINBAD}.
Using S-band travelling wave structures, it will be capable of producing sub-femtosecond electron bunches with energies of up to \SI{100}{\mega\electronvolt} and an arrival time stability of \SI{10}{\femto\second}~\cite{IPACARES,ZhuArrivalTime}.
These bunches will be optimized for injection into novel accelerating structures, such as dielectric laser acceleration (DLA) structures~\cite{EAACDLAMayet,EAACDLAKuropka} and plasma cells for laser-driven plasma wakefield acceleration (LPWA)~\cite{IPACWeikum}.

\begin{figure*}
\includegraphics[width=\textwidth]{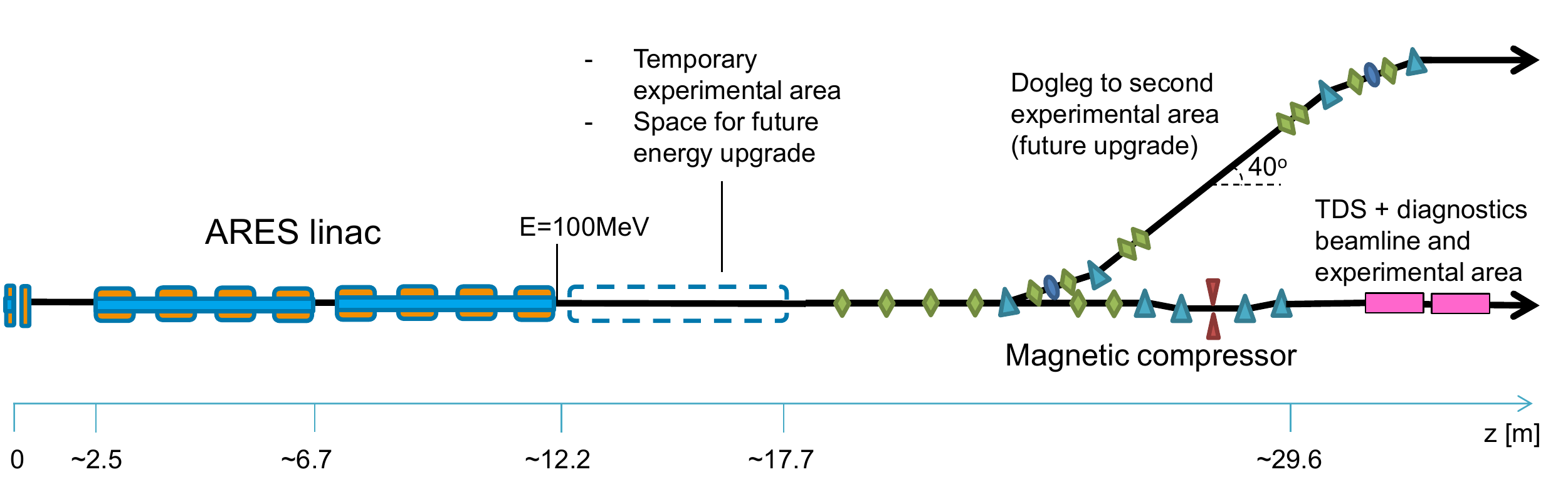}
\caption{Schematic of ARES linac showing the approximate position of the TDS cavities on the far right~\cite{IPACARES}.}
\label{fig:ARESschematic}
\end{figure*}

In order to characterize the bunches after the linac, a novel X-band transverse deflecting structure (TDS) with variable polarization will be installed.
This new design for a TDS is based on X-Band RF technology developed at CLIC~\cite{CLICGrudiev} and allows bunches to be streaked at any transverse angle within the cavity, which opens up the possibility of new measurement techniques.
The cavity will be manufactured at PSI using a high-precision tuning-free manufacturing procedure~\cite{FELEllenberger}.
Several cavities will be produced for a number of facilities at DESY and for SwissFEL, with installation of the cavity at ARES expected in 2019-2020~\cite{IPACTDS}.
The design of the TDS at ARES will consist of two 0.8-metre cavities operating at a frequency of \SI{11.9916}{\giga\hertz}, which together will be able to provide a combined transverse deflection voltage of over \SI{40}{\mega\volt}.
These cavities will be installed at the end of the magnetic compressor chicane, as shown in the schematic of the ARES linac in Fig.~\ref{fig:ARESschematic}.

\begin{figure}
\includegraphics[width=0.5\textwidth]{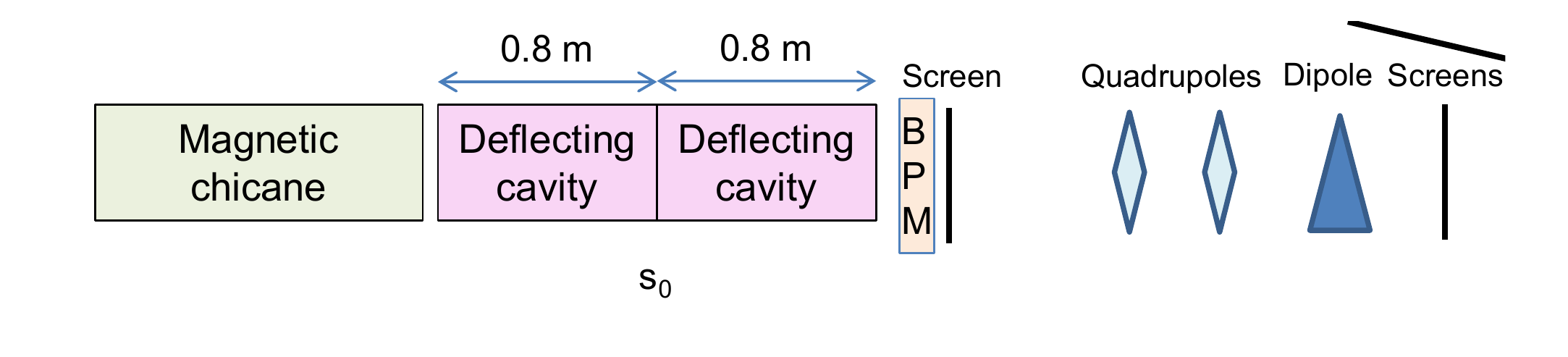}
\caption{Schematic of the lattice design at the TDS. Element positions are not yet fixed.}
\label{fig:latticeschematic}
\end{figure}

This X-band TDS will allow sub-femtosecond bunch lengths to be measured at ARES.
In addition, by combining the TDS with a dipole, a measurement can be made of the slice energy of the bunch~\cite{PhysRevSTABRoehrs,IOPlattice}.
The variable polarization also permits some measurement techniques that are not possible with a conventional TDS.
A method has been proposed to reconstruct the charge density distribution of bunches by streaking them at a number of different angles and combining the projected profiles at a screen using tomographic reconstruction techniques~\cite{IOPReconstruction}.
Additionally, by scanning over the magnetic field strength of quadrupoles placed downstream of the TDS, the transverse phase space distributions can be reconstructed.
Figure~\ref{fig:latticeschematic} shows a schematic for the lattice design of the TDS section, including screens, quadrupoles and a dipole.

These measurements are particularly challenging for the working points at ARES due to the short bunch lengths (down to sub-fs) and the relatively low energy of the beam (up to \SI{100}{\mega\electronvolt}), which means space charge effects are often non-negligible and the induced energy spread in the cavities is significant for the voltage kick required to accurately resolve the bunches.
In this paper, some of these challenges are discussed with reference to the reconstruction of the longitudinal phase space.

There are many possible applications for this measurement of the longitudinal phase space at ARES.
It could be useful in the context of DLA experiments, for example to measure microbunching structures~\cite{EAACDLAMayet}.
It could also be used to measure the energy distribution in a bunch before injection into a plasma cell.

The structure of this paper is as follows: in the following section, the theory of transverse deflection is introduced and theoretical limitations are discussed. 
Section~\ref{workingpoints} discusses the various working points under consideration and the feasibility of reconstructing the longitudinal phase space given these limitations.
In Sec.~\ref{simulations}, simulation results are presented to illustrate the technique and finally Sec.~\ref{conclusion} includes some concluding remarks and plans for further studies.

\section{Theory}
\label{theory}

A TDS imparts a time-dependent transverse kick on an electron bunch.
The rms spot size on a screen located at longitudinal position $s_1$ is therefore a combination of the spot size on the screen when the TDS is switched off, $\sigma_y^\textrm{off}$, and a contribution due to the TDS streaking:

\begin{equation}\label{Sy}
\sigma_y (s_1) = \sqrt{(\sigma_y^\textrm{off})^2+(Sc\sigma_t)^2}.
\end{equation}
Here, $\sigma_t$ is the rms bunch length at the TDS in seconds, $c$ is the speed of light and $S$ is the shear parameter, defined as

\begin{equation}
\label{shearparam}
S = M_{1,2}^y \frac{2\pi f e V_0}{c^2|p|},
\end{equation}
where ${M^y}$ is the vertical transfer matrix from the centre of the TDS to the screen, $f$ and $V_0$ are the cavity frequency and peak voltage respectively, and $p$ is the mean momentum~\cite{PhysRevSTABRoehrs}.
It has been assumed that the TDS is operated at zero-crossing.
The shear parameter relates the shift in position of a particle on the screen, $\Delta y$, to its position along the bunch relative to the central particle, $\zeta$~\cite{PhysRevSTABRoehrs}:

\begin{equation}\label{shearparameter}
\Delta y \approx S \zeta.
\end{equation}
In general,

\begin{equation}
 M_{1,2}^y = \sqrt{\beta_y(s_0)\beta_y(s_1)} \sin{\Delta\phi_y},
\end{equation}
where $\beta_y(s_0)$ and $\beta_y(s_1)$ are the $\beta$-functions at the centre of the TDS and the screen respectively, and $\Delta\phi_y$ is the phase advance between these two locations.
The temporal resolution is defined here as the bunch length in seconds when the two terms on the right-hand side of Eq.~\eqref{Sy} are equal, and can therefore be expressed as

\begin{equation}\label{Rt}
R_\textrm{t} \approx \frac{\sqrt{\epsilon_y}}{\sqrt{\beta_y (s_0)}|\sin(\Delta\phi_y)|} \frac{|p|c}{2\pi f e V_0},
\end{equation}
where $\epsilon_y$ is the vertical geometric emittance of the bunch.

\begin{table*}
\caption{\label{table:WPparameters} Working point bunch parameters at chicane exit. Working points (WP) 1 to 4 are presented in~\cite{ZhuThesis}.}
\begin{center}
\begin{tabular}{|cccccc|}
\hline
& WP1 & WP2 & WP3 & WP4 & WP5 \\
\hline
          Energy [\SI{}{\mega\electronvolt}]                           &   100.5         & 101.8         & 99.6         & 100.9                             & 84.2   \\ 
          Bunch charge [\SI{}{\pico\coulomb}]                         &   0.8           & 5.8            & 30.0            & 17.3                             & 3.0        \\
          Peak current [\SI{}{\kilo\ampere}]                            &   2.1             & 1.4            & 0.4       & 1.4                                    & 0.2         \\
          Bunch length (rms) [\SI{}{\femto\second}]                &   0.5          & 1.5            & 31.8         & 10.7                                 & 5.2   \\ 
          Rel. energy spread (rms)                                          &   \num{1.9e-3} & \num{2.4e-3} & \num{1.3e-2}   & \num{2.2e-3} & \num{2.2e-3} \\
          Energy chirp [\SI{}{\per\metre}]                               &  \num{1.1e4} & \num{4.7e3}   & \num{-6.1e2}  & \num{4.7e2}     &   \num{1.4e3}  \\
          Norm. ver. emittance [\SI{}{\milli\metre\milli\radian}]  & 0.09             & 0.29              & 1.01               & 0.32                   &  0.19  \\
\hline
\end{tabular}
\end{center}
\end{table*}

In order to reconstruct the slice bunch energy distribution, a TDS can be combined with a dipole deflecting in the other axis (defined here as the $x$ axis).
The theoretical resolution achievable for a dipole spectrometer depends on the horizontal geometric emittance ($\epsilon_x$), horizontal $\beta$-function ($\beta_x$) and dispersion ($D$) at the screen, according to the relation~\cite{BehrensGerth}

\begin{equation} \label{Eres}
R_\delta \approx \frac{\sqrt{\epsilon_x\beta_x (s_1)}}{D(s_1)}.
\end{equation}
This corresponds to the energy spread when the beam size due to dispersion and energy spread, $D(s_1) R_\delta$, is equal to the natural rms beam size, $\sqrt{\epsilon_x \beta_x(s_1)}$.

The symplectic matrix of a thick-lens cavity is given by~\cite{PRABCornacchiaEmma}

\begin{equation}
\matr{T} = 
\begin{pmatrix}
1 & L & KL/2 & 0\\ 0 & 1 & K & 0\\ 0 & 0 & 1 & 0\\ K & KL/2 & K^{2}L/6 & 1
\end{pmatrix}
\end{equation}
for a cavity with length $L$ and wavenumber $k$, where

\begin{equation}
K = \frac{eV_0 k}{pc}.
\end{equation}
From this matrix, it can be seen that the TDS induces a linear energy chirp given by

\begin{equation}
\delta(\zeta) \approx \frac{1}{6} K^2 L\zeta.
\label{corenergyspread}
\end{equation}
There are many contributions to the uncorrelated induced energy spread.
In the limit $L \to 0$, the dominant term is

\begin{equation}
\sigma_\delta = K \sigma_y\left(s_0\right).
\label{uncorenergyspread}
\end{equation}
As this term is just a function of $K$ and $\sigma_y$, it can be calculated easily and, in principle, can also be measured experimentally by scanning over the voltage.
There are other terms involved for a longer TDS, which may, in combination, be non-negligible for the TDS at ARES.
Estimating these terms would, however, be very challenging.
Using the shear parameter defined in Eq.~\eqref{shearparam}, it can be shown that~\cite{BehrensGerth}

\begin{equation}
\label{reslimit}
R_t \sigma_\delta  > \frac{\epsilon_y}{c \sin(\Delta \phi_y)}.
\end{equation}
This important result places an intrinsic limit on the product of the uncorrelated energy spread and the longitudinal resolution.
It shows that a better longitudinal resolution can only be obtained when a larger energy spread is induced in the TDS.

\section{Consideration of working points}
\label{workingpoints}

There are many working points under consideration for operation of the ARES linac, with bunch charge ranging from sub-pC to around \SI{50}{\pico\coulomb} and bunch lengths down to sub-fs scales.
Beam parameters at the chicane exit for several working points obtained from start-to-end simulations of the linac are shown in Table~\ref{table:WPparameters}~\cite{ZhuThesis}.

Using the inequality in~\eqref{reslimit}, a criterion can be found to determine whether the longitudinal phase space can be reconstructed accurately.
If the induced uncorrelated energy spread in the TDS, $\sigma_\delta$, exceeds the correlated energy spread of the input bunch, the measurement will be dominated by the energy perturbation in the TDS and so it becomes difficult to retrieve the initial longitudinal phase space distribution.

Assuming a linear energy chirp along the bunch, $h$, the initial rms correlated energy spread can be expressed as $hc\sigma_t$, where $\sigma_t$ is the initial rms bunch length in seconds.
Imposing the requirement that this rms correlated energy spread must be greater than $\sigma_\delta$, we arrive at the condition

\begin{equation}
R_t^\textrm{ref} > \frac{\epsilon_y}{|h|c^2 \sigma_t}.
\end{equation}
This gives a reference longitudinal resolution, which can be compared to the rms bunch length for an indication of the ease of reconstruction.
If this reference resolution is much shorter than the rms bunch length, an accurate reconstruction of the initial phase space should generally be possible, whereas if it is much longer, it is unlikely that the initial phase space can be reconstructed accurately.

Using this inequality, it can be shown that reconstructing the longitudinal phase space of WP1 and WP2 accurately would be extremely challenging.
In these cases, it would probably be better to turn off the field in the TDS and perform a simple spectrometer measurement of the total bunch rms energy spread instead. 
WP5 is an example that is approaching the limit for accurate reconstruction and has been used for simulations presented in the following section.

\section{Simulations}
\label{simulations}

Simulations were carried out in \emph{elegant}~\cite{elegant} using WP5 in Table~\ref{table:WPparameters}.
They do not include space charge and use a 1D CSR model.
The transverse beam parameters at the TDS entrance are shown in Table~\ref{table:BunchProperties}.
The bunch is first streaked in the vertical direction in the TDS, and then in the horizontal direction in the dipole.
The TDS is operated at zero-crossing.
In order to minimize the induced energy spread in the TDS, which is proportional to the voltage as shown in Eq.~\eqref{uncorenergyspread}, a voltage kick of just \SI{2}{\mega\volt} per cavity (i.e. \SI{4}{\mega\volt} in total) is used.
This was found to be a good compromise between induced energy spread and longitudinal resolution.

\begin{table}
\caption{\label{table:BunchProperties} Transverse statistical properties at TDS entrance for WP5.}
\begin{center}
\begin{tabular}{|cc|}
\hline
          $\sigma_{x/y}$ [\SI{}{\micro\metre}]              &   87.9 / 96.8  \\ 
          $\epsilon_{x/y}^\textrm{norm}$ [mm mrad]     &   0.224 / 0.190   \\ 
          $\beta_{x/y}$ [m]                                           &    5.69 / 8.14        \\
          $\alpha_{x/y}$                                                &    -0.377 / 0.258       \\
\hline
\end{tabular}
\end{center}
\end{table}

The setup consists of two quadrupoles for matching purposes, a dipole and a screen placed perpendicular to the beam leaving the dipole.
The dipole is assumed to be a long rectangular magnet.
The beam enters perpendicularly near the edge and leaves through the side of the magnet.
Assuming a field strength of \SI{1.4}{\tesla}, the arc length through the dipole is \SI{0.268}{\metre} and the bending angle is \SI{1.331}{\radian}.

\begin{figure}[b!]
\centering
\includegraphics[width=0.45\textwidth]{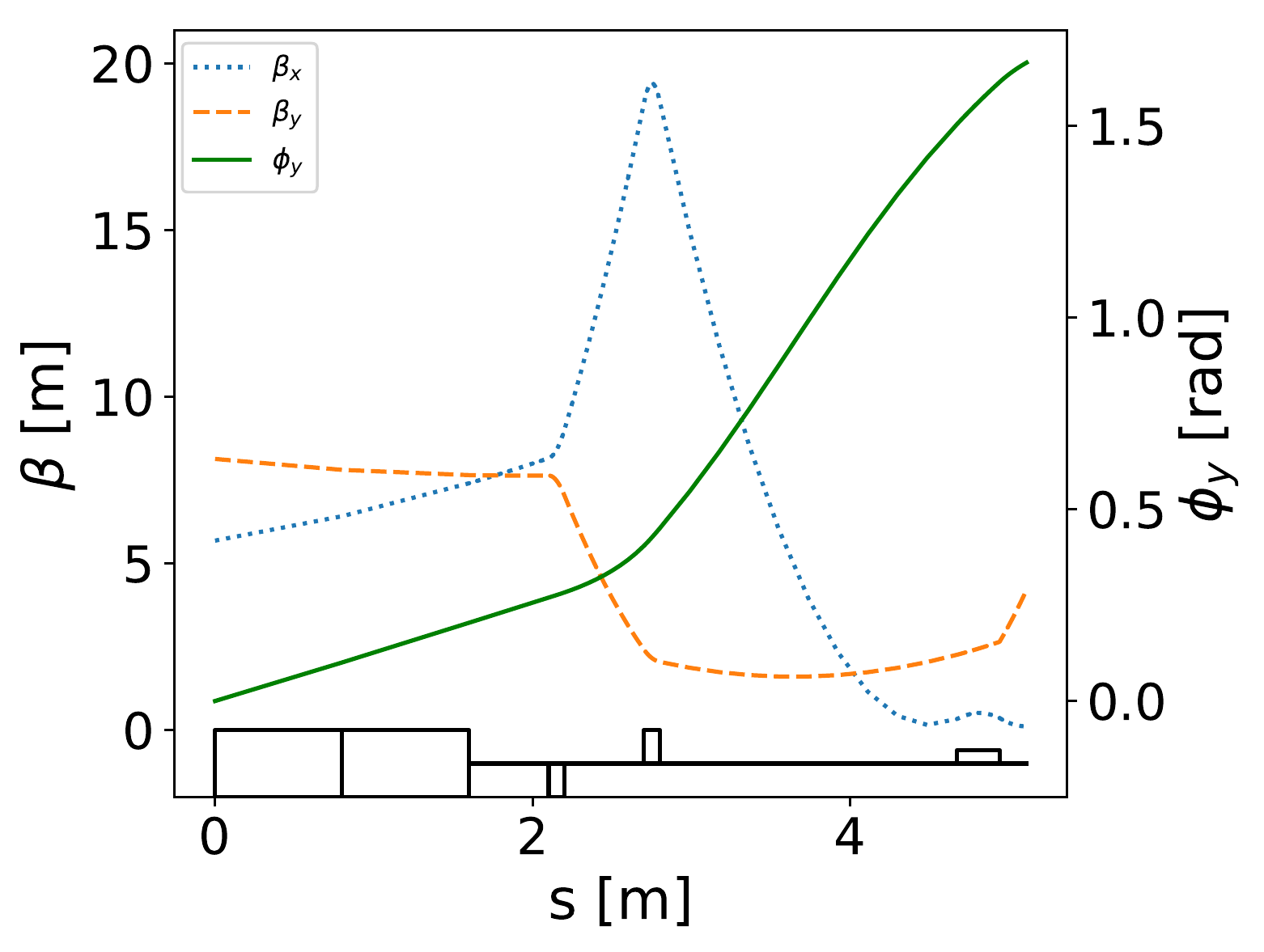}
\caption{$\beta$-functions and vertical phase advance for the lattice design, ending at the screen at $s=\SI{5.11}{\metre}$. The positions of the two TDS cavities, the two quadrupoles and the dipole are indicated at the bottom.}
\label{fig:Lattice}
\end{figure}

There are several constraints in the matching process.
In the $y$ direction, a phase shift of around $\pi/2$ is desired to optimize the longitudinal resolution, as can be seen in Eq.~\eqref{Rt}.
In the $x$ direction, a large dispersion and small $\beta_x$ at the screen is sought to optimize the energy resolution (from Eq.~\eqref{Eres}).
The final constraint is that the transverse beam size must be a sensible size on the screen.
Figure~\ref{fig:Lattice} shows the lattice used, which has been matched with respect to these conditions.
The screen is placed \SI{0.17}{\metre} after the dipole exit, which was set as the minimum possible distance for technical reasons.
Using these settings, the theoretical value for the longitudinal resolution is \SI{3.4}{\femto\second}, whereas the theoretical energy resolution ($\Delta E/E_0$) is \num{4e-5}.

\begin{figure}
\begin{subfigure}{0.45\textwidth}
\includegraphics[width=\textwidth]{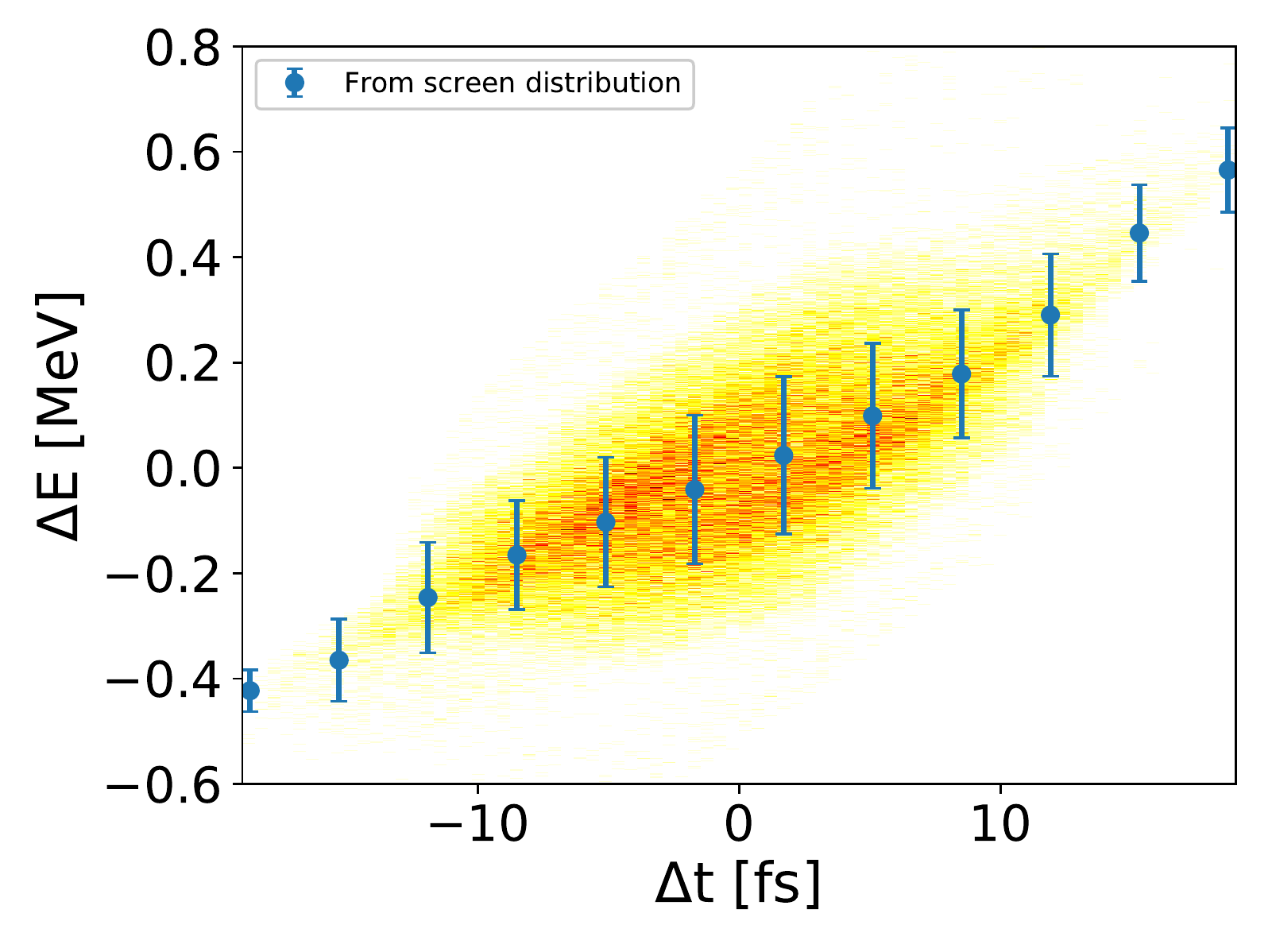}
\caption{Screen image converted to longitudinal phase space with weighted means shown as superposed dots and weighted standard deviations shown as error bars.}
\label{fig:Measured}
\end{subfigure}
\begin{subfigure}{0.45\textwidth}
\includegraphics[width=\textwidth]{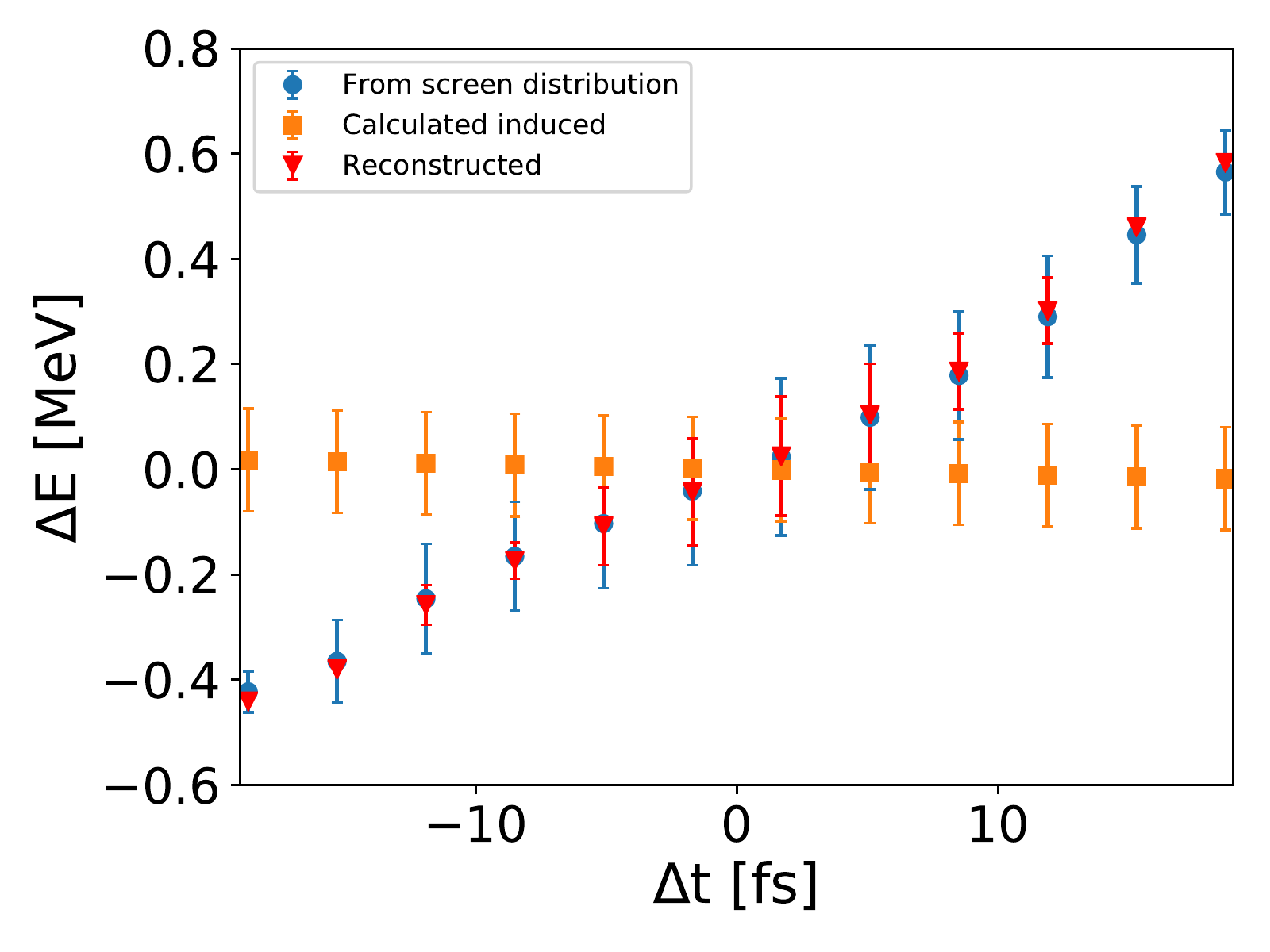}
\caption{Calculated means and standard deviations from the figure above are shown. The calculated induced energy spread is subtracted to give the reconstructed points.}
\label{fig:ErrorBars}
\end{subfigure}
\begin{subfigure}{0.45\textwidth}
\includegraphics[width=\textwidth]{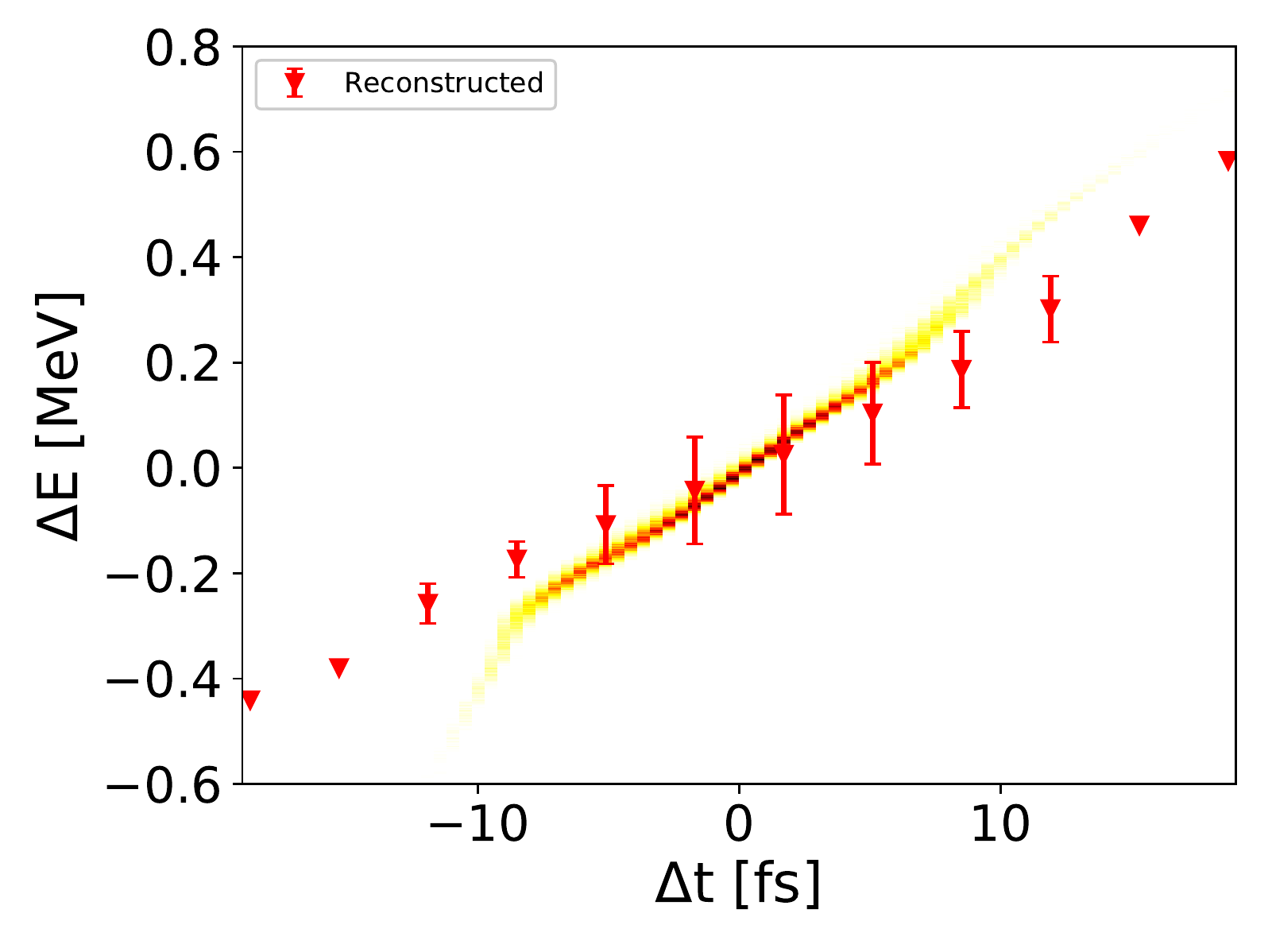}
\caption{Longitudinal phase space of the input beam with superposed error bars of the reconstruction.}
\label{fig:Reconstruction}
\end{subfigure}
\caption{Longitudinal phase space reconstruction simulation.}
\end{figure}

Figure~\ref{fig:Measured} shows the simulated screen image as a density profile.
The $y$ coordinate at the screen has been converted to arrival time ($\Delta t$) by means of the shear parameter, as given by Eq.~\eqref{shearparameter}, which has been obtained by a simulated phase scan at the cavity.
The $x$ coordinate has been converted to energy deviation ($\Delta E$) by Taylor expanding to first order the following relation obtained by geometrical arguments:

\begin{equation}
\Delta x_s = \left[d_0 + \left(\rho_1 \sin{\theta_1} - \rho_0 \sin{\theta_0} \right) \frac{\sin{\theta_1}}{\sin(\theta_0-\theta_1)} \right] \tan(\theta_0-\theta_1),
\end{equation}
where $\Delta x_s$ is the $x$ coordinate at the screen with respect to the reference trajectory, $d_0$ is the distance between the exit of the dipole and the screen, $\rho_0$ and $\rho_1$ are the reference and offset bending radii respectively, and $\theta_0$ and $\theta_1$ are the reference and offset bending angles respectively.
The weighted mean and standard deviation of this distribution are shown as error bars, with spacing given by the theoretical longitudinal resolution.

The induced energy spread, calculated using Eq.~\eqref{corenergyspread} and \eqref{uncorenergyspread}, is shown in Fig.~\ref{fig:ErrorBars} as squares with error bars.
The total standard deviation is given by

\begin{equation}
\sigma_\delta^\textrm{tot} = \sqrt{\left(\sigma_\delta^\textrm{orig}\right)^2 + \left(\sigma_\delta^\textrm{ind}\right)^2 },
\end{equation}
where $\sigma_\delta^\textrm{orig}$ is the original rms energy spread and $\sigma_\delta^\textrm{ind}$ is the induced rms energy spread.
Using this equation, the original rms energy spread can be recovered, which is shown on the plot.

Figure~\ref{fig:Reconstruction} shows the original phase space distribution with the reconstruction superposed.
The discrepancy in the reconstruction can be largely attributed to the propagation of the bunch between the TDS and the screen.
As a low voltage kick is used to limit the induced energy spread in the TDS, the longitudinal resolution is low and there is considerable mixing of the bunch slices, which alters the longitudinal phase space distribution that is reconstructed from the screen image.
This affects both the mean slice energies and the standard deviations.
The bunch used in this simulation is approaching the limit of what can be accurately reproduced, however it can be seen that useful information about the bunch properties can be extracted nevertheless.

\section{Conclusion}
\label{conclusion}
A novel X-band TDS with variable polarization is currently being designed for installation on several beamlines, including the ARES linac at the SINBAD facility at DESY.
This structure will allow measurements of time-dependent bunch properties, including the energy distribution discussed here.
By combining the TDS with a dipole and matching appropriately with quadrupoles, the longitudinal phase space of the bunch can be reconstructed in a single-shot measurement.
This measurement is challenging due to the energy spread induced in the TDS.
There is an intrinsic limit on the minimum of the product of the longitudinal resolution and the induced uncorrelated energy spread, which arises from the Panofsky-Wenzel theorem.
Therefore, for some of the working points considered for the ARES beamline, an accurate reconstruction of the longitudinal phase space is not viable, and a standard spectrometer measurement with the TDS off should be carried out instead.

A simulation of this measurement and subsequent longitudinal phase space reconstruction has been presented here for an ARES bunch.
To limit the induced energy spread to acceptable levels, the combined voltage kick has been reduced to \SI{4}{\mega\volt}.
The reconstruction agrees with the input distribution well enough to make it a useful measurement, and the discrepancies can be explained by the low temporal resolution and induced energy spread.

Further work will involve integrating collective effects into the simulations, particularly for the higher charge/current working points.
In addition, further studies are needed to combine this measurement with others mentioned in the introduction to characterize the 6D phase space of electron bunches.

\section*{Acknowledgements}
The authors would like to thank J.~Zhu for providing the input beam distributions and for many informative discussions.
Thanks are also due to W.~Kuropka and F.~Mayet for helpful discussions.


\bibliography{\jobname}

\end{document}